\documentstyle[pre,aps,epsf,graphicx]{revtex}
\begin{document}

\draft

\title{Specific heat at constant volume in the thermodynamic model}

\author{C. B. Das$^1$, S. Das Gupta$^1$ and A. Z. Mekjian$^2$}

\address{$^1$Physics Department, McGill University,
Montr{\'e}al, Canada H3A 2T8}
\address{$^2$Department of Physics and Astronomy,
Rutgers University, Piscataway, New Jersey 08855}

\date{\today}

\maketitle

\begin{abstract}
A thermodynamic model for multifragmentation which is frequently used
appears to give very different values for specific heat at constant 
volume depending upon whether canonical or grand canonical ensemble
is used.  The cause for this discrepancy is analysed.
\end{abstract}

\pacs{25.70.-z,25.75.Ld,25.10.Lx}

The motivation for this work is just one puzzle.  A thermodynamic
model, often used for fitting data, appears to give very different
answers for specific heat at constant volume depending upon whether 
the canonical or 
the grand canonical ensemble is used.  We wish to resolve this issue.
The relevant papers are \cite{Dasgupta} and \cite{Bugaev}.
Although for practical applications a much more sophisticated
two component version of the model is used, \cite{Bhattacharya,Tsang,Majumder}
here, as in \cite{Dasgupta,Bugaev} we use one kind of particle
as in the original formulation \cite{Dasgupta}.  If a system has
$A$ nucleons in a volume $V$ at temperature $T$ the system can
break up into various composites which have a binding energy
consisting of volume and surface energies.  Excited states of the
composite can also be included. Assuming that the break up takes place
only according to availability of phase space, 
it was shown in \cite{Dasgupta}
that this problem for $A$ nucleons can be solved numerically
with arbitrary accuracy very easily.  One
can also solve the problem in a grand canonical ensemble 
\cite{Bugaev,Dasgupta2}.  We do not give any details here
as they are given in many places including \cite{Dasgupta,Bugaev}.

By specific heat we will always mean specific heat per particle.
It was shown in the original paper \cite{Dasgupta} that for a 
fixed $A$, the
total number of particles and a fixed $V$, the specific heat 
as a function of temperature went through a maximum
at a certain temperature which was labelled the boiling temperature.
A numerical calculation which is rather easy and can be made sufficiently
accurate showed that for a fixed $\rho\equiv A/V$ if we increase the
number of particles the height of the peak rises and the width 
decreases.  Since one sees no reason why this behaviour should change 
at certain high value of $A$, it was concluded that the specific heat 
behaves like a delta function in the limit $A\rightarrow \infty$.
There is a physics picture one can associate with this.  Below the boiling
temparature there is a blob of liquid. 
Just above the boiling temperature this blob
changes into a system of smaller composites and nucleons.  
Qualitatively a delta
function would emerge if a finite fraction rather than an infinitesimal
fraction of the blob converts into gas with an infinitesimal increase
of temperature.

In Fig. 1 we show the specific heats in the canonical model for
$\rho=\rho_0/2.7$ for $A=200$ and $A=2000$.  All calculations shown
here will be at this $\rho$.  The grand canonical
results are also shown in the figure.  For grand canonical we solve 
\begin{eqnarray}
\rho=\sum_{k=1}^{k_m}k\exp(k\beta\mu)\tilde{\omega}_k,
\end{eqnarray}
where $k_m$ is the number of nucleons in the largest cluster
allowed in the system. Here 
\begin{eqnarray}
\tilde{\omega}_k&=&\frac{(2\pi mTk)^{3/2}}{h^3}
\exp(a_vk-\sigma(T)k^{3/2}+kT^2/\epsilon_0), \ {\rm for} \ k>1 \nonumber \\
&=& \frac{(2\pi mT)^{3/2}}{h^3}, \ {\rm for} \ k=1 .
\end{eqnarray}

To do the grand canonical calculation for $A$=2000 we set $k_m$ 
in the above equation at
2000.  The average value of $<n_k>$ is then given by,
\begin{eqnarray}
<n_k>=\exp(k\beta\mu) \tilde{\omega}_k\times V
\end{eqnarray}
where $V$ is the appropriate value
of 2000 nucleons at freeze-out, i.e., $V=2000\times 2.7/\rho_0$.  
Denoting $<n_k>$ as the average
number of composites which has $k$ nucleons (i.e., the average number
of trimers is $<n_3>$) we have 
\begin{equation}
\sum_{k=1}^{2000}k<n_k>=2000.
\end{equation}  
It should be realised
that all $A$'s (upto $\infty$) are included in the grand canonical ensemble
but setting $k_m$=2000 in eq. (1) signifies that the largest cluster has 2000 nucleons.  In terms of canonical partition functions one has
\begin{equation}
Q_{gr.can}=\sum_{K=0}^{\infty}\exp(\beta\mu K)Q_{K,k_m}
\end{equation}
where
$Q_{K,k_m}$ is the canonical partition function of $K$ nucleons but 
with the restriction that the 
largest cluster has only $k_m (=2000)$ nucleons.  The quantity $\beta\mu$ is known
from solving eq.(1) with $k_m$=2000.  Similar arguments hold when $k_m$=200
but we will not discuss this case and concentrate only on $A = k_m = 2000$.

Fig.1 shows that the specific heats in the canonical and grand canonical
models are very different.  In both the models, the peak value of the
specific heat increases when we go from 200 to 2000 particles and the
widths decrease but the results are much more dramatic in the canonical
model.  Since $A$=2000 is a large number in the context
of nuclear physics, we try to understand the cause of this difference.
In particular it is not obvious that the specific heat in the grand canonical
model will attain extraordinary heights and/or miniscule widths.  
We will show that the cause
of discrepancy between the canonical and grand canonical models is the very
large fluctuation in the particle number in the grand canonical ensemble.
We can investigate this in two ways, one more detailed than the other.
We note that
\begin{eqnarray}
Q_{gr.can}=\exp(\sum_{k=1}^{2000}V\tilde{\omega}_ke^{k\beta\mu})
\end{eqnarray}
and we can calculate fluctuation exploiting the well-known relation
\begin{eqnarray}
\frac{1}{\beta^2}\frac{\partial^2lnQ_{gr.can}}{\partial^2\mu} &=& <A^2>-<A>^2 \nonumber \\
&=& \sum_{k=1}^{2000} k^2 <n_k> 
\end{eqnarray}
But we can also exploit the fact that we know $Q_{K,k_m}$ upto rather
large values of $K$ and 
the value of $\beta\mu$ from the grand canonical calculation. 
Hence we can use the equation (5) also to calculate fluctuation.
For practical reasons, the upper limit of $K$ will have to be cut off.
The upper limit of $K$ in the sum above was 10,000.  Since we are investigating
$A$=2000 one might {\it a priori} assume this should be adequate.

The fluctuations calculated from eqs.(5) and (7) are shown in Fig. 2.
One sees that there is a temperature above which the fluctuations are small.
At these temperatures the grand canonical value of specific heat is
indistinguishable from the canonical value.  But as temperature is lowered,
fluctuations grow rapidly and the results begin to diverge.

It is interesting to study fluctuations further.  The probability of
$K$ particles being in the grand canonical ensemble is 
$\propto e^{K\beta\mu+lnQ_K}$ (Eq. 5) and we plot in Fig.3 
$\exp[\beta\mu(K-A)+lnQ_K-lnQ_A]$.  This takes the value 1 at $K=A$
and in the normal picture of grand canonical
ensemble, would drop off rapidly
on either side of $A$.  This does happen at temperature higher than the boiling
temperature.  The case at temperature T=7.7 MeV corresponds to a standard
scenario.  But the situation at temperature 7.3 MeV is drastically different.
The probability does not maximise at $K=A$ but at a lower value.  
It is also very spread out with a periodic structure.  At 
temperature 2.0 MeV, the probability of having no particle is higher than the
probability of having $K=A$.  We notice that here also there is a periodicity
in the probability distribution.  The periodicity is 2000 and is 
linked with the fact that in the case studied the largest composite has
2000 nucleons and at low temperatures, this composite will play
a significant role. 

We can now understand why the specific heat curve is so flat in the grand 
canonical ensemble in Fig 1.  Even though the average number $A$ is 2000,
the ensemble contains large components of $K<A$ (thus have lower density
and peaks of specific heat below 7.1
MeV) and $K>A$ (which have peaks at higher than 7.1 MeV).  It is this smearing
which makes the specific heat peak much lower and much wider.

In Fig. 4 we have shown canonical and grand canonical results for the
total energy of 2000 particles.  The canonical result suggests that
starting from a low temperature, energy increases at a finite rate 
(implies a finite value of $C_V$), followed by a sudden rise (will 
lead to infinite specific heat), followed again by a regular 
behaviour.
Thus the transition region is marked by two different values of $C_V$
with a delta function switched in between.  In the grand canonical 
model for 2000 particles one sees only the discontinuity in the value
of $C_V$.  If one was calculating the $C_V$ directly in either the
canonical or grand canonical model and for a very large system there
is indeed a delta function in $C_V$, one will miss the delta function
since it has zero width.  To see that there is indeed one, we should 
instead calculate the total energy and see that at a given
temperature $T$ there is a huge difference in the total
energy for an infinitesimal increase in $T$.

One might get the impression that for finite systems the use of a grand 
canonical ensemble is very dangerous.  For many observables it is quite
acceptable.  However, a recent work shows \cite{Das} that interpreting
data according to grand canonical ensemble can lead to a serious error
in estimating temperature.  It thus depends on the particular observable
being calculated.

Returning briefly to the interesting periodicity seen in Fig.3, it arises
because at low temperatures it is advantageous for the system to form
large clusters.  In the example studied in Fig. 3 where the largest cluster 
size was 2000, at temperature 2 MeV when the system has 8000 nucleons we have
essentially 4 clusters, each of size 2000 (we are confining ourselves to
canonical calculation of course). This becomes less precise
at higher temperature.  For example at 6 MeV temperature if the system
has 4100 nucleons nearly 4000 of them are distributed in large clusters
(more precisely, on the average, 3950.22 nucleons are bound in clusters
of sizes between 1800 and 2000) and nearly 100 of them are in lighter 
clusters (149.78 on the average).  If we now go to a system of
6100 nucleons, on the average 5940.22 are bound in very heavy clusters and
159.78 in lighter clusters.  Approximately speaking, at low temperatures,
$lnQ_{K+2000}\approx lnQ_{K}+ln(V\tilde{\omega}_{2000})$.

The low temperature periodic structure at $T = 2$ MeV can be qualitatively
understood using the following results. As $T \rightarrow 0$, the system 
will go to the largest cluster allowed, and in this case, $k_m = 2000$.
For example, at $K = 10,000$ (the total number of nucleons in the system)
a result of 5 clusters of size 2000 follows. The mean number of cluster
of size $k_m (=2000)$ is:
\begin{eqnarray}
<n_{k_m}> &=& \omega_{k_m} \frac{Q_{K-{k_m}}}{Q_K} \nonumber 
\end{eqnarray}
Also the factorial moment $<n_{k_m}(n_{k_m}-1)>$ is given by,
\begin{eqnarray}
<n_{k_m}(n_{k_m}-1)> &=& {(\omega_{k_m})}^2 \left(\frac{Q_{K-2{k_m}}}{Q_K}\right). \nonumber 
\end{eqnarray}
And in general,
\begin{eqnarray}
<n_{k_m}(n_{k_m}-1)...(n_{k_m}-n+1)> &=& {(\omega_{k_m})}^n \left(\frac{Q_{K-n{k_m}}}{Q_K}\right).
\end{eqnarray}
Thus the points at 8000, 6000, 4000, 2000 and 0 can be related to the
factorial moments of the $n_{k_m}$ distributions as $T \rightarrow 0$.
At $T = 0$ these are $5 \times 4 \times 3 \times 2 \times 1, 5 \times 4 \times 3 \times 2, 5 \times 4 \times 3, 5 \times 4$ and 5 at $K = 0, 2000, 4000, 6000$ and at 8000 respectively.
The heights of the peaks in Fig. 3 will be determined by,
\begin{center}
\begin{math}
\frac{1}{{\omega_{_m}}^n}<n_{k_m}(n_{k_m}-1)...(n_{k_m}-n+1)>\frac{Q_K}{Q_{k_m}}e^{\beta \mu (K-(n+1)k_m)}
\end{math}
\end{center}
as $T \rightarrow 0$. Even at $T =2$ MeV, this expression is very accurate.

We acknowledge communications from K. A. Bugaev and I. Mishustin.
This work is supported in part by the Natural Sciences and Engineering
Research Council of Canada, Fonds Nature et Technologies of Quebec 
and the U.S. Department of Energy Grant No. DE FG02-96ER40987.

\begin{figure}
\epsfxsize=5.5in
\epsfysize=7.0in
\centerline{\rotatebox{270}{\epsffile{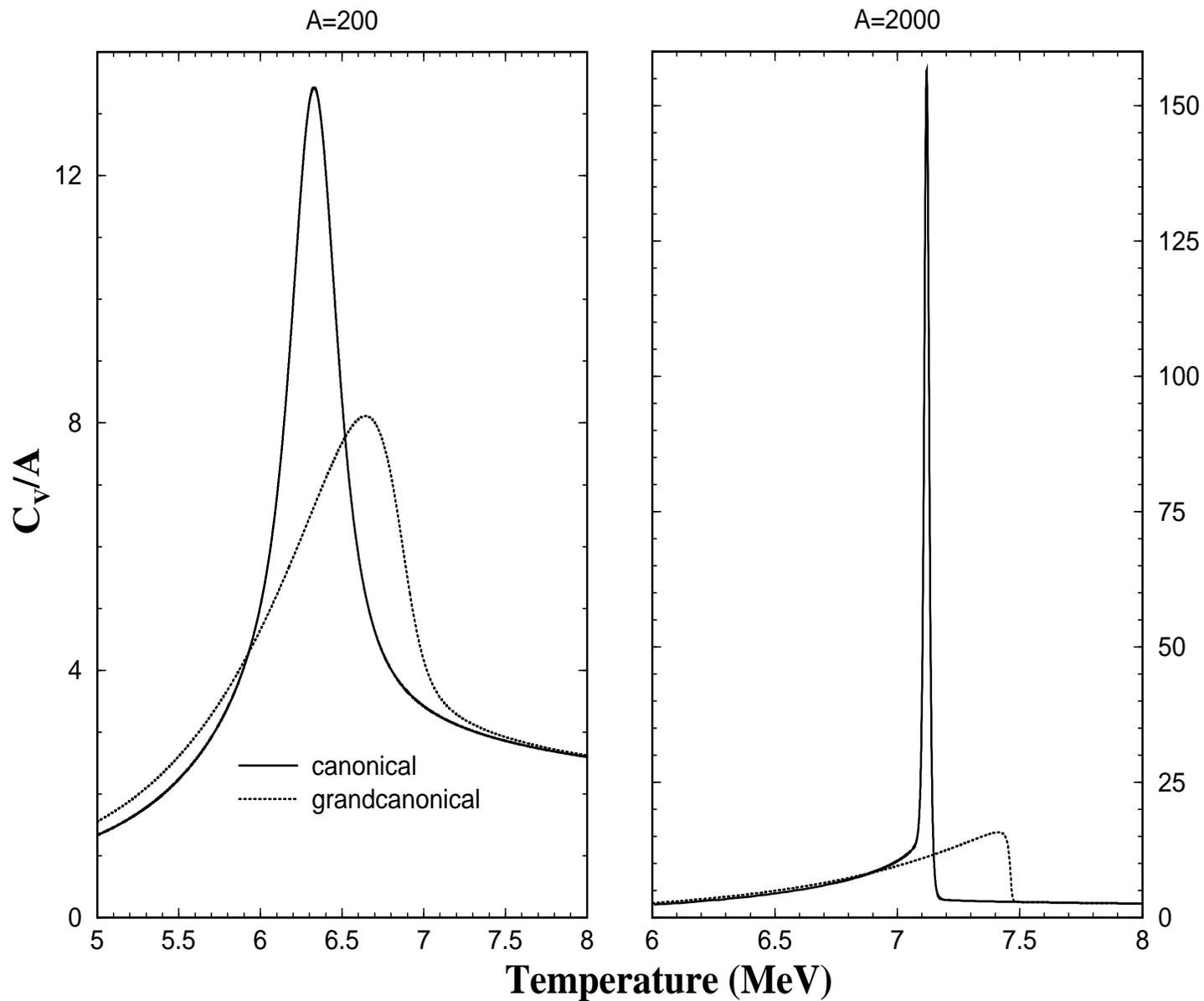}}}
\vskip 0.5in
\caption{Specific heat per particle at constant volume when the system has
total number of particles 200 and 2000.  Canonical and grand canonical
values are shown.}
\end{figure}

\begin{figure}
\epsfxsize=5.5in
\epsfysize=7.0in
\centerline{\epsffile{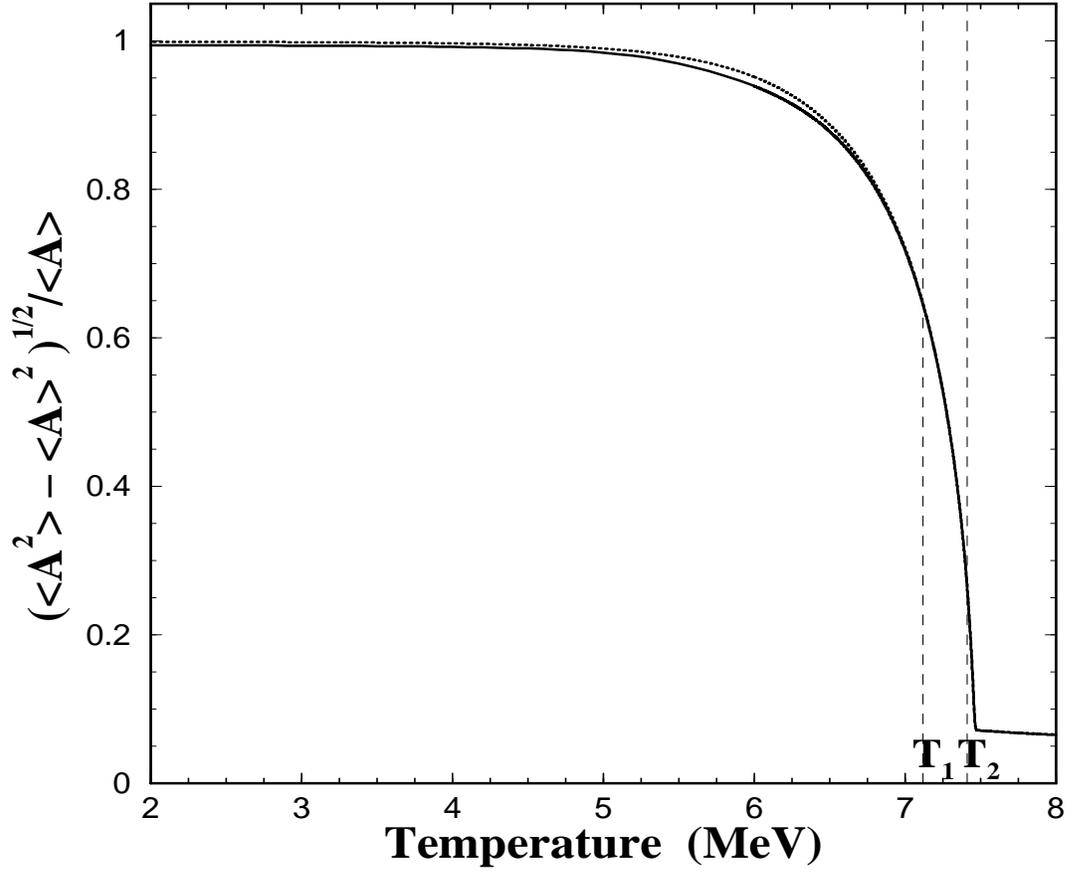}}
\caption{Fluctuations calculated using eqs.(5) and (7). 
The solid line corresponds 
to using eq. (5) with $K$ cut off at 10,000 and the dotted line
corresponds to using eq. (7). $T_1$ correspond to 
the temperature where the specific heat maximises in the canonical 
calculation and $T_2$ to the temperature of highest specific heat 
in the grand canonical calculation.}
\end{figure}

\begin{figure}
\epsfxsize=5.5in
\epsfysize=7.0in
\centerline{\rotatebox{270}{\epsffile{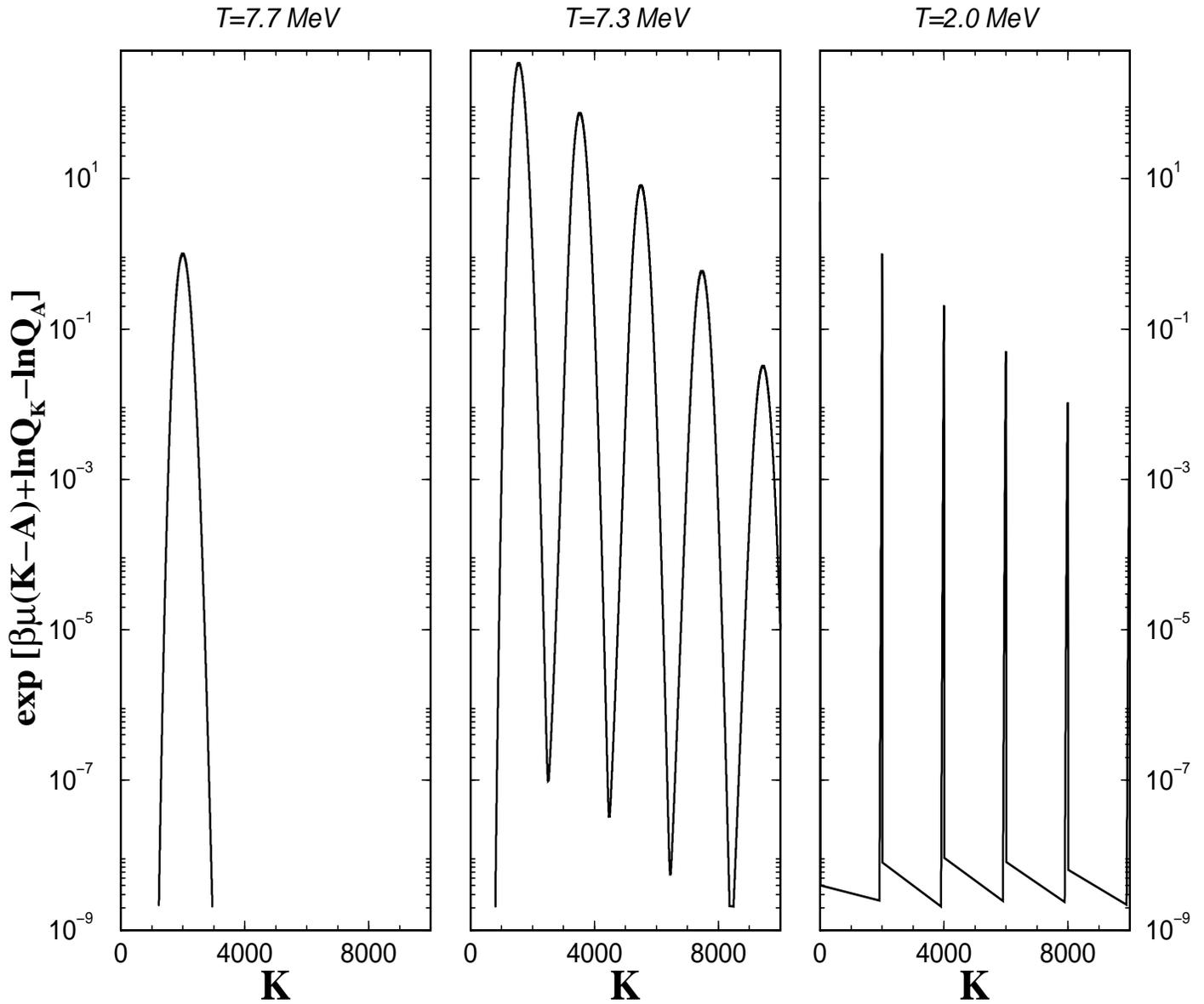}}}
\vskip 0.5in
\caption{These graphs show the spread of particle numbers in the grand
canonical ensemble when the average particle number is 2000.  The
spread is very narrow at temperature 7.7 MeV but becomes quite
wide at lower temperatures.}
\end{figure}

\begin{figure}
\epsfxsize=5.5in
\epsfysize=7.0in
\centerline{\epsffile{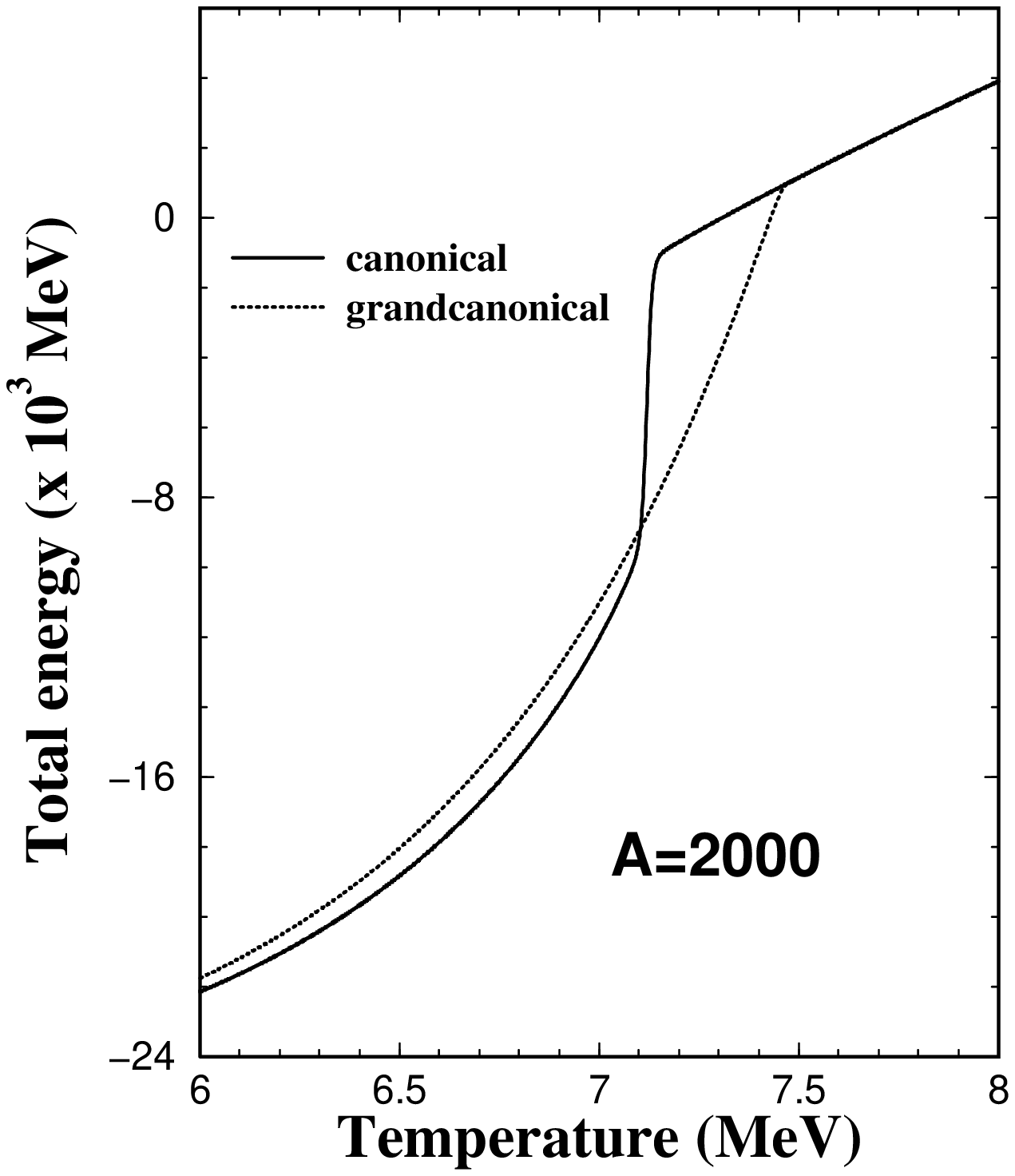}}
\caption{Caloric curves in the canonical and grandcanonical models, for
a system of 2000 particles.}
\end{figure}


\begin{references}
\bibitem{Dasgupta} S. Das Gupta and A. Z. Mekjian, Phys. Rev C{\bf
57}, 1361 (1998).

\bibitem{Bugaev} K. A. Bugaev, M. I. Gorenstein, I. N. Mishustin and
W. Greiner, Phys. Rev C{\bf 62}, 044320 (2000).

\bibitem{Bhattacharya} P. Bhattacharyya, S. Das Gupta and 
A. Z. Mekjian, Phys. Rev C{\bf 60}, 054616 (1999).

\bibitem{Tsang} M. B. Tsang {\it et al}, Phys. Rev. C{\bf 64}, 054615
(2001).

\bibitem{Majumder} A. Majumder and S. Das Gupta, Phys. Rev. C{\bf
61}, 034603 (2000).

\bibitem{Dasgupta2} S. Das Gupta and A. Z. Mekjian, Phys. Rep. {\bf 
71}, 131 (1981).

\bibitem{Das} C.B.Das, S. Das Gupta, X.D. Liu and M.B. Tsang,
Phys. Rev. C{\bf 64}, 044608, (2001).

\end{references}
\end{document}